\documentclass[12pt]{iopart}

\usepackage{iopams}
\usepackage{subfigure}
\usepackage{graphicx}

\begin{document}
\title[Improved surface quality of anisotropically etched silicon \{111\} planes.]{Improved surface quality of anisotropically etched silicon \{111\} planes for mm-scale integrated optics.}

\author{J P Cotter$^1$\footnote{Present address: University of Vienna, Faculty of Physics, VCQ, Boltzmanngasse 5, A-1090 Vienna, Austria}, I Zeimpekis$^2$, M Kraft$^2$\footnote{Present address: University of Duisberg-Essen, Faculty of Engineering Sciences, Bismarckstr. 81, D-47057 Duisburg, Germany} and E A Hinds$^1$}

\address{$^1$ The Centre for Cold Matter, The Blackett Laboratory, Imperial College London, SW7 2AZ, UK.}
\address{$^2$ Nano research group, Electronics and Computer Science, University of Southampton, Highfield, Southampton, SO17 1BJ, UK}
\ead{joseph.cotter@univie.ac.at}
\ead{ed.hinds@imperial.ac.uk}

\begin{abstract}
We have studied the surface quality of millimeter-scale optical mirrors produced by etching CZ and FZ silicon wafers in potassium hydroxide to expose the  $\{111\}$ planes. We find that the FZ surfaces have four times lower noise power at spatial frequencies up to $500\,\mbox{mm}^{-1}$. We conclude that mirrors made using FZ wafers have higher optical quality.
\end{abstract}

\maketitle

\section{Introduction}

Potassium Hydroxide (KOH) is often used as an anisotropic etchant to expose the $\{111\}$ planes of silicon\,\cite{seidel90,seidel90_2}. These planes can act as highly reflecting mirrors, able to steer light around a device. Because they need not be parallel or perpendicular to the body of the wafer such mirrors provide a simple way to couple light from sources outside the chip into the plane of an integrated device.  A wide range of devices are produced in this way, for applications ranging from solar cells\,\cite{Jianhua02,munoz09,yoon09} to integrated optics\,\cite{sadler97,mori05,perney06} and atom chips\,\cite{atomchipbook,pollock09,pollock11,nshii13}. 

While this method readily produces micron-sized mirrors that are excellent reflectors\,\cite{trupke06},  larger structures present more difficulty. First, the KOH etch rate is typically  $1\,\mu$m/min\,\cite{williams03}, corresponding to days of etching for millimeter-sized structures. Second, such deep etches leave the exposed \{111\} planes with significant roughness\,\cite{pollock09,holke99}. The low spatial-frequency components change the shape of the surface, altering the intensity distribution of the reflected light. The high-frequency (sub-micron) roughness reduces the reflectivity of the surface.  A final stage of isotropic etching is sometimes employed to improve the surface quality. Wet etching in a solution of hydrofluoric, nitric and acetic acids (HNA) can smooth the surface\,\cite{schwartz76}, but changes its shape\,\cite{pollock09}. Isotropic plasma etching can flatten the surface, but introduces roughness at short length scales\cite{laliotis12}. It is therefore useful to seek an improved deep etching method that may not require further polishing.

So far, the silicon wafers used to make mm-scale $\{111\}$-plane mirrors\,\cite{pollock09} have been from Czochalski zone (CZ) crystals. Since precipitates and complexes of oxygen play an important role in determining the quality of surfaces etched using KOH \cite{holke99}, and since float zone (FZ) silicon is essentially free of oxygen\,\cite{fransilla_book}, we wondered whether FZ wafers would produce better mirrors.

In this paper, we compare the roughness of $\{111\}$ mirror surfaces etched in CZ and FZ silicon wafers. In section \ref{sec:Fabrication} we discuss the fabrication method used to create the mm-scale $\{111\}$ silicon mirrors. In Sec.\,\ref{sec:surface} we present images of the mirror surfaces etched in the two types of wafer. A white-light interferometer provides quantitative measurements including noise power spectra. Our findings are summarised in Sec.\,\ref{sec:summary}. 
 
\section{Fabrication}
\label{sec:Fabrication}

Both silicon wafers were 3mm thick, cut on the $[100]$ plane. The CZ wafers were $10\,$cm-diameter, boron-doped p-type, with a resistivity of approximately $10\,\Omega\,$cm. The FZ wafers were phosphorus-doped n-type ($3\,\mbox{k}\Omega\,$cm) with $15\,$cm diameter.  We cleaned any organic or ionic contamination off the wafer surfaces using the RCA (Radio Corporation of America) process, then deposited $200\,$nm of low-stress silicon nitride on both sides by low pressure chemical vapour deposition.  The cleaning ensured nitride layers with strong adhesion to the silicon and free of pinholes.  On one side of each wafer, a $1\,\mu$m-thick layer of positive photoresist (S1813) was patterned with square apertures by uv exposure. Through these, we opened squares in the silicon nitride by reactive ion etching with CHF$_{3}$,  in preparation for wet-etching square pyramids into the silicon.

A solution of potassium hydroxide (KOH) was made by dissolving reagent grade pellets in $14\,l$ of de-ionised water to a concentration of 37\,wt.\%, as measured by a specific gravity hydrometer. This was held in a temperature-controlled bath at $80^{\circ}$C, and the concentration was maintained using a re-condensing lid. Polytetraflouroethylene (PTFE) holders suspended both wafers together in the etching bath, and also protected the back faces from any unwanted etching. Oxygen was bubbled through the solution to discourage the build up of hydrogen on the surface of the wafer, which can affect surface morphology\,\cite{palik91,campbell95}. The pyramids that formed were etched to give an $8.2\,$mm opening - large enough to allow our microscope access to the exposed $\{111\}$ planes.  The depth of the etch was $1.9$\,mm, which took $47\,$hours for the CZ wafer and $43\,$hours for the FZ wafer. The etch rate should not be affected by the low boron and phosphorous doping\,\cite{seidel90_2}. We attribute the difference to the higher purity of the FZ wafer.

\begin{figure*}[b]
\centering
\subfigure[CZ mircosope images]{\includegraphics[width= 0.45 \columnwidth]{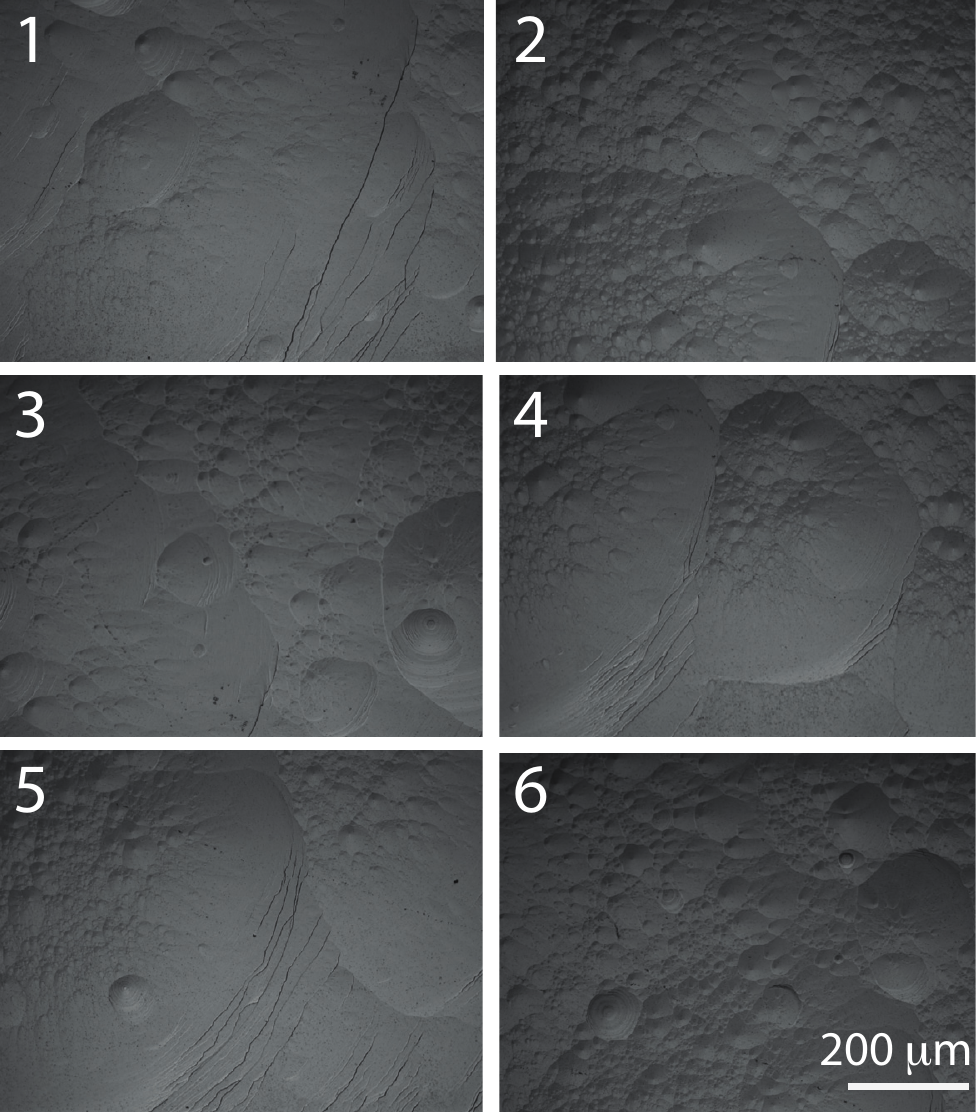}}
\hspace{1cm}
\subfigure[FZ mircosope images]{\includegraphics[width= 0.45 \columnwidth]{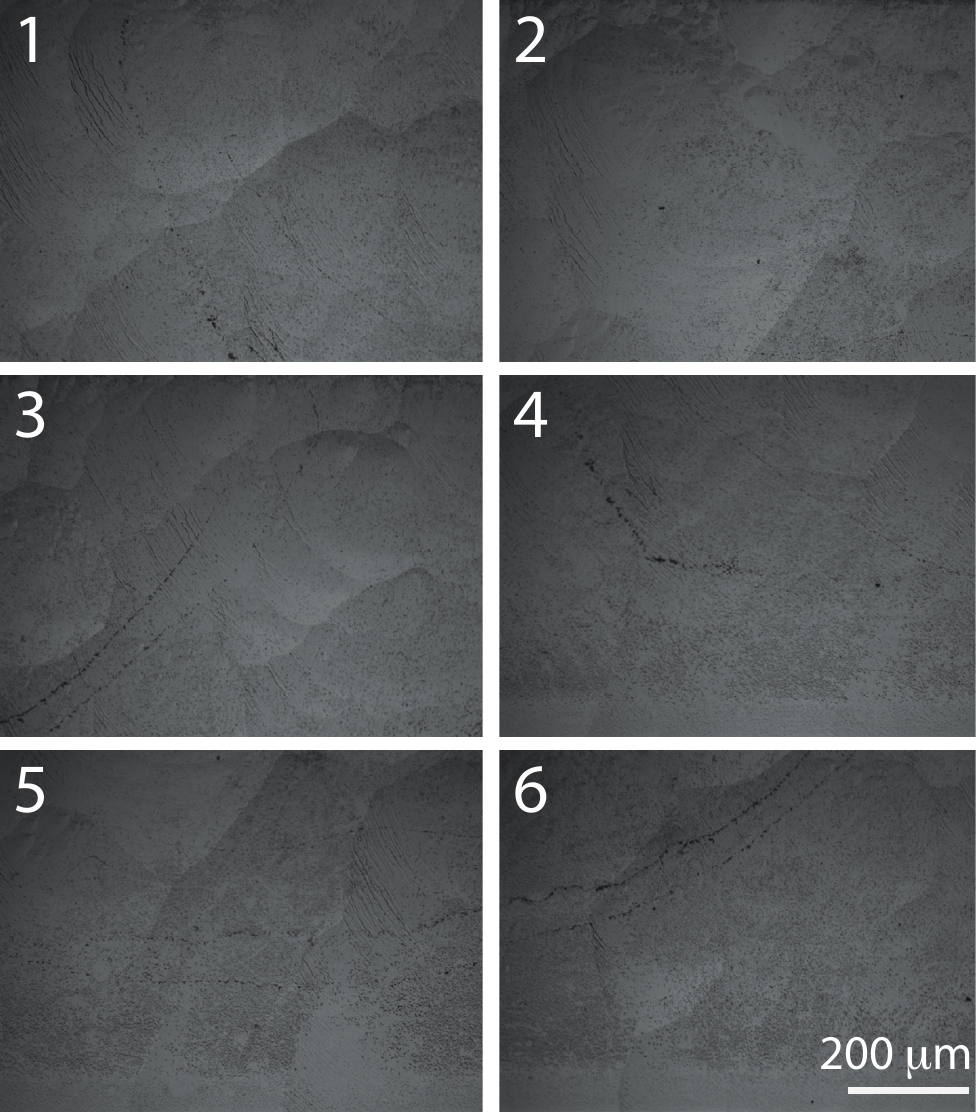}}
\caption{Microscope images of etched \{111\} planes. (a) Images of six regions on the CZ wafer. These show roughness over a wide range of length scales, particularly evident in the $10-100\,\mu$m range and particularly in panels $2$ and $6$. (b) Six regions on the FZ wafer.  These are smoother. The deposit of particulates, seen as small black spots are iron contamination from the KOH solution. Once cleaned from the surface they do not affect the roughness.
\label{fig:Micro}}
\end{figure*}

\section{Surface characteristics}
\label{sec:surface}

On initial inspection under the microscope, the FZ and CZ wafers both appeared to be coated in a deposit of small particles, aproximately $1\,\mu$m in size, appearing as black spots. A $5\,$minute immersion in fuming nitric acid had no impact on these particles, indicating that they were not organic contaminants. Energy dispersive X-ray spectroscopy found that they were iron, with the same density of contamination on both wafers. It seems most probable that the iron deposits come from the etching solution, as described in \cite{kwa95,tanaka00}, and could have been avoided, or at least reduced, by using KOH pellets of higher purity. The spots were cleaned by a brief immersion in a solution of hydrocloric acid and hydrogen peroxide.

\subsection{Microscope images}

Fig. \ref{fig:Micro}(a) shows typical microscope images taken from six different areas on an exposed $\{111\}$ face of the CZ  wafer. The side of the image is parallel to the side of the square pyramid opening. Representative images from the FZ wafer are shown in Fig. \ref{fig:Micro}(b). The main difference is immediately obvious: the bubbly, ``orange peel" texture in the size range $10-100\,\mu$m - particularly clear in panels (2) and (6) - is much reduced in the CZ images. It is not possible to say more about the comparative roughness from these optical microscope images because they are only qualitative.

\begin{figure*}[b]
\centering
\subfigure[CZ interferometer images]{\includegraphics[width=0.45\columnwidth]{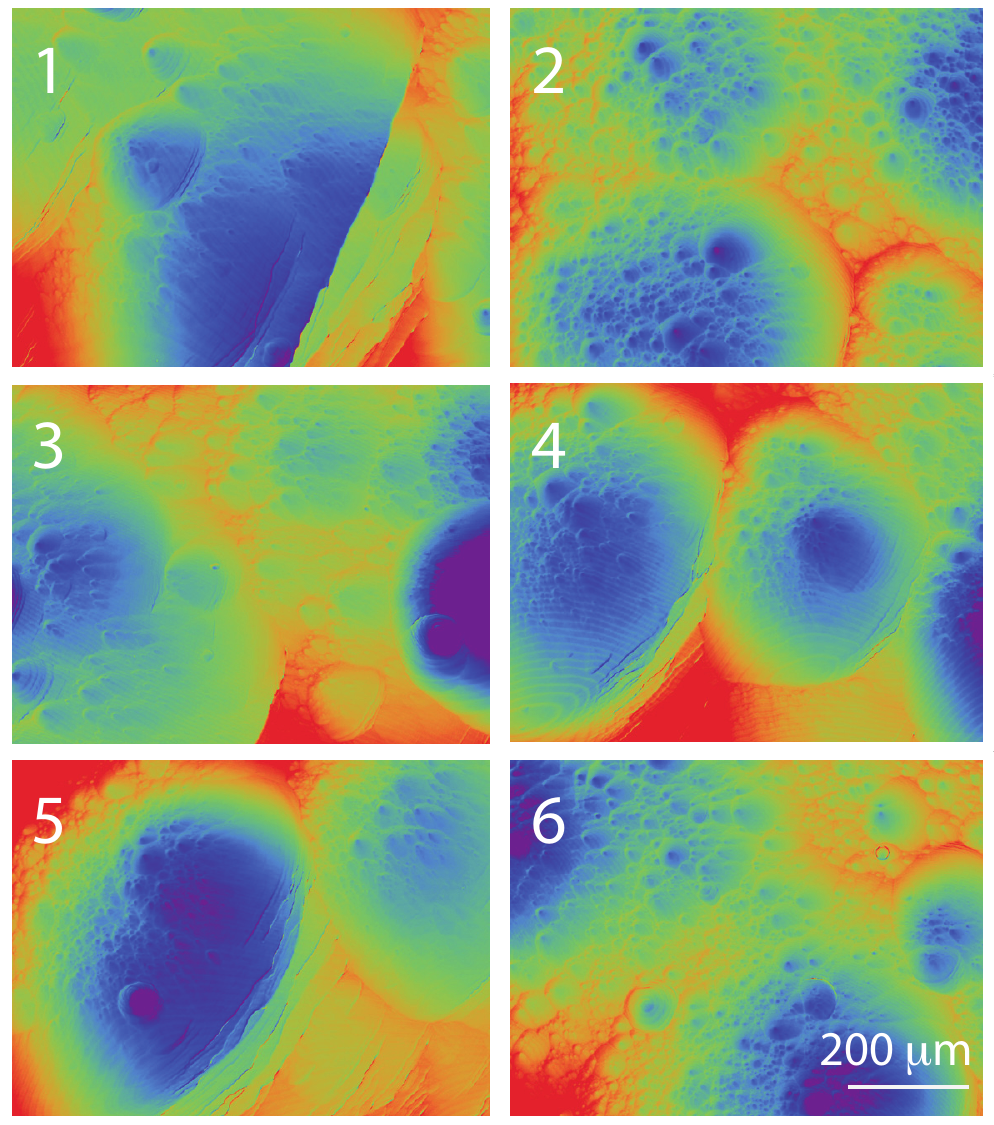}}
\hspace{1cm}
\subfigure[FZ interferometer images]{\includegraphics[width=0.45\columnwidth]{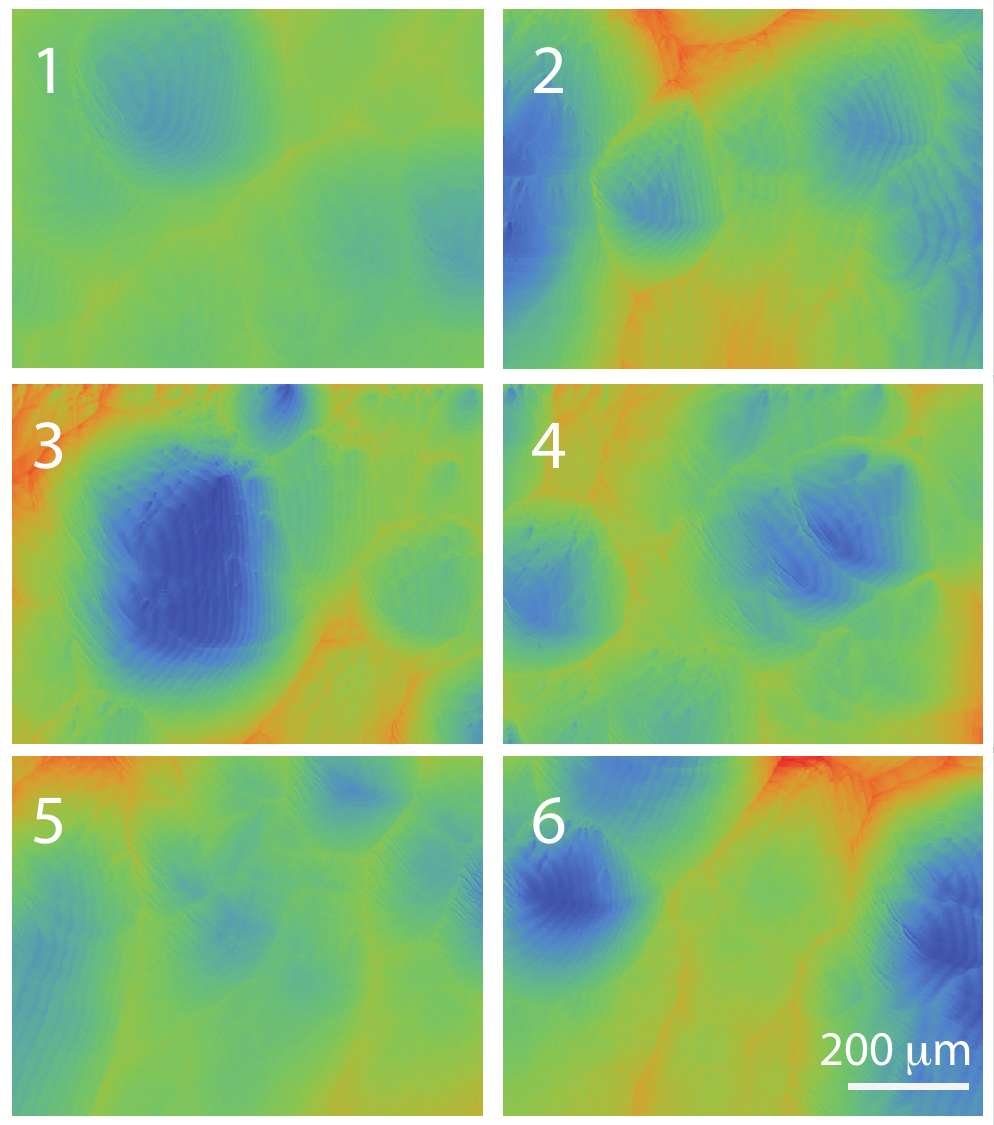}}
\caption{
White-light interferometer images of etched \{111\} planes taken over the same regions as Fig.\,\ref{fig:Micro}. (a)  CZ wafer. (b) FZ wafer,  plotted on the same colour scale as (a). One sees that the height variation is significantly less in (b) than it is in (a).
\label{fig:Int}}
\end{figure*}

\subsection{White-light interferometry}

\begin{figure}[t]
\centering
\includegraphics[width=0.7\columnwidth]{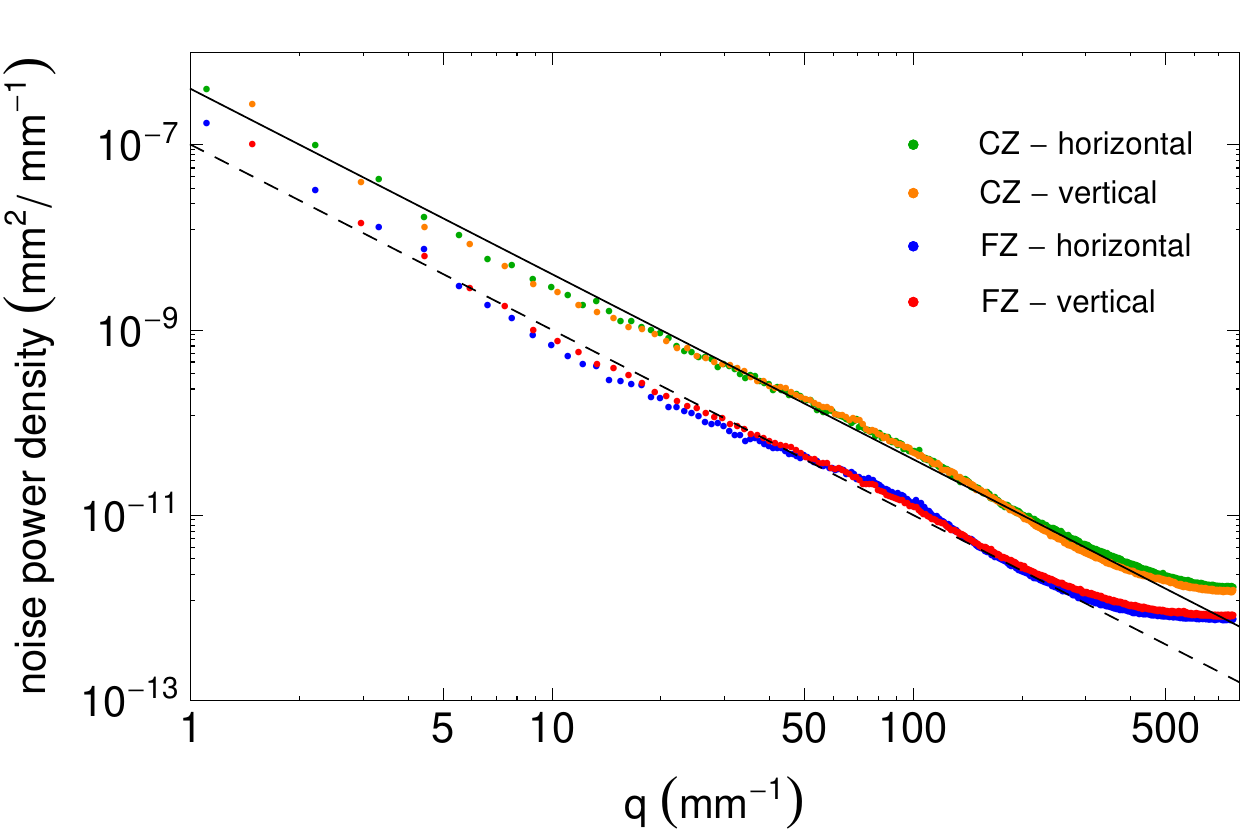}
\caption{Spectral power density of the height noise in the mirror surfaces, derived from the interferograms shown in Fig.\,\ref{fig:Int}. The wavenumber $q$ is the inverse of the wavelength. Upper data points: discrete power spectra for the CZ mirror measured along horizontal (green) and vertical (orange) lines. Lower data points: FZ noise power spectra for horizontal noise power in the horizontal (blue) and vertical (red) scans. Solid line: The model spectrum $4\times 10^{-7} /q^2$. Dashed line: Model spectrum $1\times 10^{-7} /q^2$.
\label{fig:spectrum}}
\end{figure}

Fig.\,\ref{fig:Int} shows the same twelve regions as Fig.\,\ref{fig:Micro}, imaged using a white-light interferometer, with plane sloping  backgrounds subtracted. As well as showing once again that the FZ surfaces have less noise, these images provide calibrated height information from which we derive noise spectra. For each horizontal row in Fig. \,\ref{fig:Int}(a) - the CZ images - we calculate the 1D noise power spectrum and average over all the rows and images to produce the set of green data points in Fig.\,\ref{fig:spectrum}. The orange points, showing the noise taken over vertical scans, are not significantly different. The solid line shows the power-law spectrum, $4\times 10^{-7} /q^2$ with $q$, being the wavenumber (1/wavelength) in mm$^{-1}$.  This simple model is in remarkably close agreement with the spectrum that we have measured until it runs into the noise floor of the interferometer (a few times $10^{-13}\,\mbox{mm}^2/\mbox{mm}^{-1}$) at approximately $500\,\mbox{mm}^{-1}$.

The lower, blue and red data points are the noise power spectra derived in the same way from the images of the FZ surface in Fig.\,\ref{fig:Int}(b). We see that the FZ mirror is systematically smoother over the whole frequency range measured, with a spectrum that is well described by the simple approximation (dashed line) $1\times 10^{-7} /q^2$. The form of the spectrum means that in both cases the roughness is dominated by the lowest frequency components. 

\section{Summary and conclusion}
\label{sec:summary}
We have studied the roughness of the mirror surfaces exposed by deep anisotropic KOH etching of silicon. We find that the height of the surface has a noise power spectrum that follows the inverse square of the frequency. On comparing mirrors etched in CZ and FZ wafers we find that the FZ mirrors have lower noise power at all frequencies up to $500\,\mbox{mm}^{-1}$ by a factor of approximately 4. We tentatively ascribe this difference to the lower level of oxygen precipitates and complexes in FZ silicon. We conclude that mirrors formed in FZ wafers will require less polishing to achieve the same optical quality as CZ mirrors and we surmise that the highest possible quality is likely to be achieved using FZ material.

\ack
This work was supported by the CEC Seventh Framework project 247687 (AQUTE), the UK EPSRC and the Royal Society.


\section*{References}
\begin{thebibliography}{10}

\bibitem{seidel90}
H.~Seidel, L.~Csepregi, A.~Heuberger, and H.~Baumgartel.
\newblock Anisotropic etching of crystalline silicon in alkaline solutions.
\newblock {\em J. Electrochem. Soc.}, 137:3612, 1990.

\bibitem{seidel90_2}
H.~Seidel, L.~Csepregi, A.~Heuberger, and H.~Baumgartel.
\newblock Anisotropic etching of crystalline silicon in alkaline solutions ii.
\newblock {\em J. Electrochem. Soc.}, 137:3626, 1990.

\bibitem{Jianhua02}
Z.~Jianhua, W.~Aihua, P.~P. Altermatt, M.~A. Green, J.~P. Rakotoniaina, and
  O.~Breitenstein.
\newblock High efficiency pert cells on n-type silicon substrates.
\newblock {\em Conference Record of the Twenty-Ninth IEEE: Photovoltaic
  Specialists Conference}, pages 218-- 221, 2002.

\bibitem{munoz09}
D.~Munoz, P.~Carreras, J.~EscarrŽ, D.~Ibarz, S.~Mart'n de~Nicol‡s, C.~Voz, J.M.
  Asensi, and J.~Bertomeu.
\newblock Optimization of {KOH} etching process to obtain textured substrates
  suitable for heterojunction solar cells fabricated by {HWCVD}.
\newblock {\em Thin Solid Films}, 517:3578--3580,, 2009.

\bibitem{yoon09}
J.~Yoon et. al.
\newblock Ultrathin silicon solar microcells for semitransparent, mechanically
  flexible and microconcentrator module designs.
\newblock {\em Nature Materials}, 7:907, 2008.

\bibitem{sadler97}
D.~J. Sadler, M.~J. Garter, C.~H. Ahn, S.~Koh, and A.~L. Cook.
\newblock Optical reflectivity of micromachined \{111\}-orientated silicon
  mirrors for optical input-output couplers.
\newblock {\em J. Mi}, 7:263, 1997.

\bibitem{mori05}
T.~Mori, S.~Sugawara, K.~Adachi, K.~Mori, and S.~Satoh.
\newblock Silicon microoptical mirrors to make close parallel beams with
  conventional laser diodes.
\newblock {\em Journal of Microelectromechanical Systems}, 14:37--43, 2005.

\bibitem{perney06}
N.~M.B. Perney, J.~J. Baumberg, M.~E. Zoorob, M.~D.~B. Charlton, S.~Mahnkopf,
  and C.~M. Netti.
\newblock Tuning localized plasmons in nanostructured substrates for surface
  enhanced {R}aman scattering.
\newblock {\em Optics Express}, 14:847--857, 2006.

\bibitem{atomchipbook}
J.~Reichel and V.~Vuletic, editors.
\newblock {\em Atom Chips}.
\newblock Wiley, 2011.

\bibitem{pollock09}
S.~Pollock, J.~P. Cotter, A.~Laliotis, and E.~A. Hinds.
\newblock Integrated magneto-optical traps on a chip using silicon pyramid
  structures.
\newblock {\em Optics Express}, 17:14109, 2009.

\bibitem{pollock11}
S.~Pollock, J.~P. Cotter, A.~Laliotis, F.~Ramirez-Martinez, and E.~A. Hinds.
\newblock Characteristics of integrated magneto-optcal traps for atom chips.
\newblock {\em New J. Phys.}, 13:043029, 2011.

\bibitem{nshii13}
C.~C. Nshii, M.~Vangeleyn, J.~P. Cotter, P.~F. Griffin, E.~A. Hinds, C.~N.
  Ironside, P.~See, A.~G. Sinclair, E.~Riis, and A.~S. Arnold.
\newblock A surface-patterned chip as a strong source of ultracold atoms for
  quantum technologies.
\newblock {\em nature nanotechnology}, 2013.

\bibitem{trupke06}
M.~Trupke, F.~Ramirez-Martinez, E.~A. Curtis, J.~P. Ashmore, S.~Eriksson, E.~A.
  Hinds, Z.~Moktadir, C.~Gollasch, M.~Kraft, G.~Vijaya Prakash, and J.~J.
  Baumberg.
\newblock Pyramidal micromirrors for microsystems and atom chips.
\newblock {\em App. Phys. Letts.}, 88:071116, 2006.

\bibitem{williams03}
K.~R. Williams, K.~Gupta, and M.~Wasilik.
\newblock Etch rates for micromachining processing-part ii.
\newblock {\em Journal of Microelectromechanical Systems}, 12:761--778, 2003.

\bibitem{holke99}
A.~Holke and H.~T. Henderson.
\newblock Ultra-deep anisotropic etching of (110) silicon.
\newblock {\em J. Micromech. Microeng.}, 9, 1999.

\bibitem{schwartz76}
B.~Schwartz and H.~Robbins.
\newblock Chemical etching of silicon, iv. etching technology.
\newblock {\em J. Electrochem. Soc.}, 123:1903--1910, 1976.

\bibitem{laliotis12}
A.~Laliotis, M.~Trupke, J.~P. Cotter, G.~Lewis, M.~Kraft, and E.~A. Hinds.
\newblock {ICP} polishing of silicon for high-quality optical resonators on a
  chip.
\newblock {\em J. Micromech. Microeng}, 22:125011, 2012.

\bibitem{fransilla_book}
Sami Franssila.
\newblock {\em Introduction to Microfabrication}.
\newblock J. Willey, 2004.

\bibitem{palik91}
E.~D. Palik, O.~J. Glembocki, I.~Heard Jr., P.~S. Burno, and L.~Tenerz.
\newblock Etching roughness for (100) silicon surfaces in aqueous {KOH}.
\newblock {\em J. Appl. Phys}, 70:3291, 1991.

\bibitem{campbell95}
S.~A. Campbell, K.~Cooper, L.~Dixon, S.~N. Port, and D.~J. Schiffrin.
\newblock Inhibition of pyramid formation in the etching of {Si}
  p$\langle100\rangle$ in aqueous potassium hydroxide-isopropanol.
\newblock {\em J. Michromech. Microeng}, 5:209--218, 1995.

\bibitem{kwa95}
T.~A. Kwa and R.~F. Wolffenbuttel.
\newblock Effect of solution contamination on etched silicon surfaces.
\newblock {\em J. Micromech. Microeng.}, 5:95--97, 1995.

\bibitem{tanaka00}
H.~Tanaka, Y.~Abe, T.~Yoneyama, J.~Ishikawa, O.~Takenaka, and K.~Inoue.
\newblock Effects of small amount of impurities on etching of silicon in
  aqueous potassium hydroxide solutions.
\newblock {\em Sensors and Actuators}, 82:279--273, 2000.

\end{thebibliography}
\end{document}